\documentclass[aps,prl,reprint,nofootinbib,nobibnotes,superscriptaddress,floatfix]{revtex4-2}
\usepackage{amsmath,amssymb,bm,graphicx}
\usepackage[colorlinks=true,citecolor=blue,linkcolor=blue,urlcolor=blue]{hyperref}
\newcommand{\Tr}{\operatorname{Tr}}
\newcommand{\EE}{\mathbb E}
\newcommand{\Pauli}{\mathcal P}
\hypersetup{hypertexnames=false}
\begin{document}
\let\SupplementAuthor\author
\let\SupplementAffiliation\affiliation
\let\SupplementEmail\email
\let\SupplementMakeTitle\maketitle

\title{Universal stabilizer correlations in chaotic eigenstates}
\author{Zhi-Han Liu}
\affiliation{Hefei National Research Center for Physical Sciences at the Microscale and School of Physical Sciences, University of Science and Technology of China, Hefei 230026, China}
\author{Hao-Yue Qi}
\email[Corresponding author: ]{qhaoyue@mail.ustc.edu.cn}
\affiliation{Hefei National Research Center for Physical Sciences at the Microscale and School of Physical Sciences, University of Science and Technology of China, Hefei 230026, China}
\author{Wei Zheng}
\email[Corresponding author: ]{zw8796@ustc.edu.cn}
\affiliation{Hefei National Research Center for Physical Sciences at the Microscale and School of Physical Sciences, University of Science and Technology of China, Hefei 230026, China}
\affiliation{CAS Center for Excellence in Quantum Information and Quantum Physics, University of Science and Technology of China, Hefei 230026, China}
\affiliation{Hefei National Laboratory, University of Science and Technology of China, Hefei 230088, China}

\begin{abstract}
Entanglement and nonstabilizerness provide complementary perspectives on quantum-state structure. In this work, we find that the stabilizer correlation, defined as the difference between the R\'enyi-$2$ mutual information and the mutual stabilizer R\'enyi entropy,    approaches a universal constant for chaotic eigenstates. This constant is determined solely by the symmetries, and quantifies the correlation that does not contribute to nonstabilizerness. We identify this universal stabilizer correlation analytically via random-state ensembles, which give $\log_2 7$ (GOE) and $\log_2 4$(GUE) without extra symmetries, $\log_2 14$ and $\log_2 8$ under spin-inversion parity or half-filling magnetization. While imposing reflection or translation symmetry leaves these values unchanged. Exact calculations for a chaotic mixed-field Ising chain confirm the predicted plateau and finite-size approach. Integrable free-fermion and strongly disordered eigenstates behave qualitatively differently. This universal stabilizer correlation reveals a characteristic structure of chaotic eigenstates beyond entanglement or nonstabilizerness alone.

\end{abstract}
\maketitle

Quantum entanglement plays a central role in quantum information and many-body physics~\cite{horodecki,amico}. It underlies teleportation, superdense coding, entanglement-based quantum key distribution, and measurement-based quantum computation~\cite{bennettteleport,bennettdense,ekert,raussendorf}. The scaling of entanglement entropy reveals criticality, while its growth after a quench probes information spreading and thermalization~\cite{amico,eisert,calabresequench}. Entanglement alone, however, does not fully characterize state structure or computational complexity. Clifford circuits can generate highly entangled stabilizer states that nevertheless remain efficiently simulable classically~\cite{gottesman,Aaronson,bravyigosset}. Nonstabilizerness quantifies the departure from stabilizer states ~\cite{bravyi}. Stabilizer R\'enyi entropies (SREs) measure nonstabilizerness through Pauli expectation values~\cite{leone} and have been used to study many-body states and phase transitions~\cite{olivieroisingsre,tarabungahaug,colluragaussian,ramirezshannon}. The SRE, however, does not resolve the nonstabilizerness associated with correlations and can obscure criticality~\cite{tarabunga2023,tarabungahaug,wangmagic}. The mutual SRE (mSRE) instead characterizes nonstabilizerness shared between subsystems~\cite{tarabunga2023,fraumutual}, revealing ground-state criticality and transitions in monitored dynamics that are less apparent in the SRE~\cite{tarabunga2023,tarabungahaug,wangmagic}.

Entanglement and nonstabilizerness also provide distinct perspectives on quantum chaos, which is intimately connected to thermalization in isolated systems~\cite{deutsch,srednicki,dalessio}. Chaotic eigenstates exhibit volume-law entanglement~\cite{kimhuse,dalessio,vidmarrigol,leblondxxz}, approaching random-state entropy benchmarks~\cite{page}. Eigenstate fluctuations reveal additional structure beyond the mean entropy~\cite{geraedts,rodrigueznieva}. Regarding nonstabilizerness, chaotic dynamics can generate substantial nonstabilizerness~\cite{leonechaos,goto,odavicquench}, and chaotic eigenstates approach the SRE of random states~\cite{turkeshi,iannottiu1}. These results show that nonstabilizerness is an essential ingredient of chaotic quantum states. However, neither entanglement nor nonstabilizerness alone fully captures their structure: entanglement does not distinguish stabilizer from nonstabilizer resources, while nonstabilizerness by itself does not resolve how this resource is distributed across correlations. This complementarity suggests that the interplay between entanglement and nonstabilizerness is central to quantum chaos.

In this work, we reveal a universal relation between the R\'enyi-$2$ mutual information and the mSRE of chaotic eigenstates. Specifically, we introduce the stabilizer correlation as the difference between the R\'enyi-$2$ mutual information and the mSRE. We find that, in the thermodynamic limit, the stabilizer correlation of chaotic eigenstates approaches a constant determined solely by the symmetries of the system. We determine this constant analytically using random-state ensembles. Without additional symmetries, the GOE and GUE yield stabilizer correlations of $\log_{2} 7$ and $\log_{2} 4$, respectively. Imposing spin-inversion parity or fixing the magnetization at half filling gives 
$\log_{2} 14$ for real states and $\log_{2} 8$ for complex states, whereas reflection or translation symmetry leaves the unconstrained value unchanged. We test these random-state predictions using a chaotic mixed-field Ising chain. In contrast, the free-fermion integrable chain exhibits a pronounced system-size dependence in its stabilizer correlation, while eigenstates in the strongly disordered chain show much weaker stabilizer correlations, well below the random-state predictions.


\begin{figure}[tbp]
 \includegraphics[width=0.96\columnwidth]{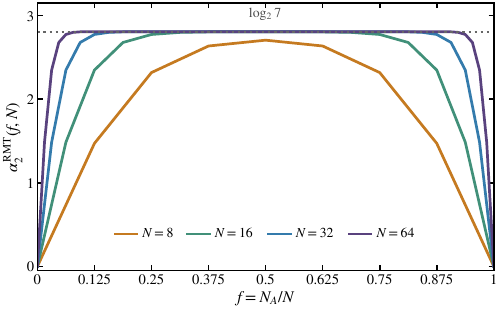}
 \caption{Random-state reference $\alpha_2^{\rm RMT}(f,N)$ in the GOE, calculated from the exact fourth-moment sums in Eq.~\eqref{eq:finite}. For every fixed $0<f<1$, the value tends to $\log_2 7$ as $N$ grows, while the empty and full partitions give zero.}
 \label{fig:random}
\end{figure}

\textit{Stabilizer correlations.---}
Consider a pure state $\hat\rho_{AB}=|\psi\rangle\langle\psi|$ of $N$ qubits, partitioned into regions $A$ and $B$ with $N_A+N_B=N$. We denote the subsystem fraction by $f=N_A/N$. The corresponding Hilbert-space dimensions are $d_A=2^{fN}$, $d_B=2^{(1-f)N}$, and $d=2^N=d_Ad_B$. For a region $\Omega\in\{A,B,AB\}$, let $\hat\rho_\Omega=\Tr_{\bar{\Omega}}\hat{\rho}_{AB}$ denote the reduced state and let $\hat P_\Omega$ run over the phase-free Pauli strings $\Pauli_{N_\Omega}=\{\hat I,\hat X,\hat Y,\hat Z\}^{\otimes N_\Omega}$. $\bar\Omega$ denotes the complementary subsystem when $\Omega=A$ or $B$. The nonstabilizerness of $\hat{\rho}_{\Omega}$ is quantified by the stabilizer R\'enyi-$2$ entropy \cite{leone},
$M_2(\hat\rho_\Omega)=-\log_2
 \frac{\sum_{\hat P_\Omega}|\Tr(\hat\rho_\Omega\hat P_\Omega)|^4}
 {\sum_{\hat P_\Omega}|\Tr(\hat\rho_\Omega\hat P_\Omega)|^2}.$
Here, mixed stabilizer states are defined as mixed states obtained by partially tracing pure stabilizer states \cite{Aaronson,leone,tarabungahaug,tarabunga2023}. 

The mSRE is defined as
\begin{equation}
 \mathcal L_2=M_2(\hat\rho_{AB})-M_2(\hat\rho_A)-M_2(\hat\rho_B).
 \label{eq:mutual}
\end{equation}
It compares the nonstabilizerness of the full state with that retained in the two reduced states, thereby characterizing the nonstabilizerness associated with their correlations. In particular, it vanishes for a product state, even when both factors have nonzero SRE.
With $S_2(\hat\rho)=-\log_2\Tr\hat\rho^2$, the Rényi-$2$ mutual information is defined as $I_2=S_2(\hat\rho_A)+S_2(\hat\rho_B)-S_2(\hat\rho_)$, which quantifies the total correlations between subsystems $A$ and $B$. For the pure state considered here, it reduces to twice the entanglement entropy, $I_2=2S_2(\hat\rho_A)$. We define the stabilizer correlation as~\cite{tarabunga2023,tarabungahaug}
\begin{align}
  \alpha_2 &= I_2- \mathcal{L}_2, \\
  &=-\log_2 \frac{\displaystyle\sum_{\hat P_A}|\Tr(\hat\rho_A\hat P_A)|^4\sum_{\hat P_B}|\Tr(\hat\rho_B\hat P_B)|^4}
{\displaystyle\sum_{\hat P}|\Tr(\hat\rho_{AB}\hat P)|^4},
 \label{eq:identity}
\end{align}
which reflects, to some extent, the part
of the correlations that does not contribute to
nonstabilizerness described by mSRE. At fixed total correlation, a larger $\alpha_2$ means less nonstabilizerness in the correlations. A stabilizer state has only stabilizer correlation, $\alpha_2=I_2$, as exemplified by a maximally entangled Bell pair. The mSRE $\mathcal{L}_2$ and stabilizer correlation $\alpha_2$ thus describe different aspects of the total correlations. In the following, we demonstrate that the stabilizer correlation approaches a universal constant for $N,N_A,N_B \gg 1$ in chaotic states, differs in the integrable examples, and is much smaller in the strongly disordered chain over the sizes studied.

\begin{table}[t]
\caption{Stabilizer correlation at fixed $0<f<1$, with each constraint imposed separately. Real and complex denote GOE and GUE. Real random states are possible only at $k=0$ and, for even $N$, $k=N/2$; complex states can occupy any momentum sector.}
\label{tab:symmetry}
\begin{ruledtabular}
\begin{tabular}{lcc}
Constraint & Real & Complex\\
\colrule
None & $\log_2 7$ & $\log_2 4$\\
Spin-inversion parity $\pm$ & $\log_2 14$ & $\log_2 8$\\
Reflection parity $\pm$ & $\log_2 7$ & $\log_2 4$\\
Momentum $k$ & $\log_2 7$ & $\log_2 4$\\
Magnetization, half filling & $\log_2 14$ & $\log_2 8$
\end{tabular}
\end{ruledtabular}
\end{table}

\textit{Stabilizer correlation of random states.---}
Consider a generic chaotic Hamiltonian with total Hilbert-space dimension $d$ that may possess symmetries. In a fixed symmetry sector of dimension $D$, we approximate the statistics of typical eigenstates using a random-state ensemble of the same dimension. We consider the Gaussian orthogonal and unitary ensembles (GOE and GUE), whose random states have real and complex components, respectively. Denote the Pauli sums in Eq.~\eqref{eq:identity} by
$F_\Omega=\sum_{\hat P_\Omega\in\Pauli_{N_\Omega}}x_{\hat P_\Omega}^4,$ with $x_{\hat P_\Omega}=\Tr(\hat\rho_\Omega\hat P_\Omega)$. Let $\EE$ denote the ensemble average. The ratio defines the random-matrix reference
\begin{equation}
 \alpha_2^{\rm RMT}=\log_2\frac{\EE F_{AB}}{(\EE F_A)(\EE F_B)}.
 \label{eq:momentreference}
\end{equation}
This reference need not equal $\alpha_2$ at finite size. Fluctuations of the Pauli sums vanish in the large-system limit, so a typical state's $\alpha_2$ approaches $\alpha_2^{\rm RMT}$. The concentration bounds and detailed moment derivations are given in the Supplemental Material (SM)~\cite{supplement}.

Without symmetry restrictions, $D=d$. In a real state, a Pauli string containing an odd number of $\hat Y$ factors has zero expectation because $\hat P_\Omega^T=-\hat P_\Omega$. Every nonidentity even-$Y$ string is symmetric, traceless, and obeys $\hat P_\Omega^2=\hat I$. Embedded in the full Hilbert space, it has $d/2$ eigenvalues of each sign. Normalizing $d$ independent real standard Gaussian amplitudes and averaging gives
$\EE x_{\hat P_\Omega}^4=\frac{12}{(d+2)(d+6)}.$ This equation holds for every nonidentity Pauli string with an
even number of $\hat Y$ factors.
There are $d_\Omega(d_\Omega+1)/2$ even-$Y$ strings, including the identity. Note that the identity contributes $x_{\hat I_\Omega}=1$, we obtain
$\EE F_\Omega=1+\frac{6(d_\Omega^2+d_\Omega-2)}{(d+2)(d+6)},
 \quad\Omega=A,B,AB.$
Substitution into Eq.~\eqref{eq:momentreference} yields the finite-size random-matrix reference
\begin{equation}
 \alpha_2^{\rm RMT}(f,N)=\log_2
 \frac{7d/(d+6)}
 {\displaystyle\prod_{\Omega=A,B}\left[1+\frac{6(d_\Omega^2+d_\Omega-2)}{(d+2)(d+6)}\right]}.
 \label{eq:finite}
\end{equation}
At fixed $0<f<1$, both subsystems grow with $N$, giving $\EE F_A,\EE F_B\to1$, whereas $\EE F_{AB}\to7$. Hence, we have
$\alpha_2^{\rm RMT}(f,N)\to\log_2 7.$ Figure~\ref{fig:random} shows this asymptotic behavior. 
The empty and full partitions instead have $\alpha_2=0$. For complex random states, all $d^2-1$ nonidentity strings contribute the fourth moment $3/[(d+1)(d+3)]$. Their full-system sum tends to $3$, and the identity adds $1$, while both subsystem sums again tend to $1$. The limiting stabilizer correlation is therefore $\log_2 4$.

\begin{figure}[tbp]
 \includegraphics[width=\columnwidth]{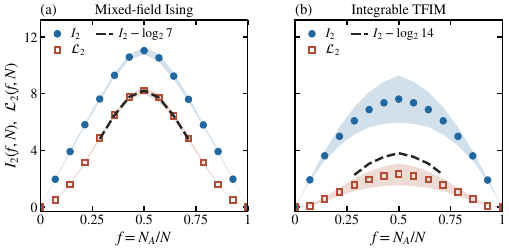}
 \caption{Mutual information and mSRE versus $f$ at $N=14$, averaged over $100$ eigenstates. Circles show $I_2$ and squares show $\mathcal L_2$. The blue and orange bands indicate the standard deviations across eigenstates. Dashed lines show $I_2-\log_2 7$ in (a) and $I_2-\log_2 14$ in (b), for $4/14\le f\le10/14$. (a) Mixed-field Ising chain. (b) TFIM with even spin-inversion parity.}
 \label{fig:entropies}
\end{figure}

\textit{Effect of symmetries.---}
Within a symmetry sector, Pauli expectations are determined by $\hat\Pi\hat P\hat\Pi$, with $\hat\Pi$ the sector projector. For spin inversion, $\hat U=\prod_i\hat X_i$ commutes with $\hat H$, and $\hat\Pi_s=(\hat I+s\hat U)/2$, $s=\pm1$, projects onto a sector of dimension $D=d/2$. A string with an odd total number of $\hat Y$ and $\hat Z$ factors anticommutes with $\hat U$, and has zero expectation. Real states additionally require even $\hat Y$ parity, leaving $(d^2+2d)/4$ strings.
Among these, $\hat I$ and $\hat U$ have fixed expectations $1$ and $s$, each contributing $1$ to $F_{AB}$. All other allowed operators are traceless within the sector and square to the identity, with $D/2$ eigenvalues of either sign. Their fourth moment is $\EE x_{\hat P_\Omega}^4=\frac{12}{(D+2)(D+6)}$ for GOE. In subsystem, only the identity $\hat I$ has a fixed expectation $1$; the other strings are evaluated as $\hat P_\Omega\otimes\hat I_{\bar\Omega}$ in the same full-system sector. Summing gives
\begin{equation}
 \begin{aligned}
 \EE F_{AB}&=\frac{14d}{d+12},\\
 \EE F_\Omega&=1+\frac{12(d_\Omega^2+2d_\Omega-4)}{(d+4)(d+12)},
 \quad\Omega=A,B.
 \end{aligned}
 \label{eq:spinmeans}
\end{equation}
Substituting these
expressions into Eq.~\eqref{eq:momentreference} yields finite-size random-matrix
reference. At fixed $0<f<1$, $\EE F_A,\EE F_B\to1$ and $\EE F_{AB}\to14$, giving $\alpha_2^{\rm RMT}\to\log_2 14$ with exponentially small corrections. For complex states, the same procedure gives $\log_2 8$. The fluctuations around these ensemble averages vanish in the large-system limit~\cite{supplement}.

For translation, reflection, and fixed magnetization, projected Pauli operators can have eigenvalues other than $\pm1$, requiring the general fourth-moment formula~\cite{supplement}. Table~\ref{tab:symmetry} lists the constants with each constraint imposed separately. Reflection or compatible momentum leaves the full-space values unchanged, whereas half filling gives $\log_2 14$ and $\log_2 8$ for real and complex states.

\begin{figure}[tbp]
 \includegraphics[width=\columnwidth]{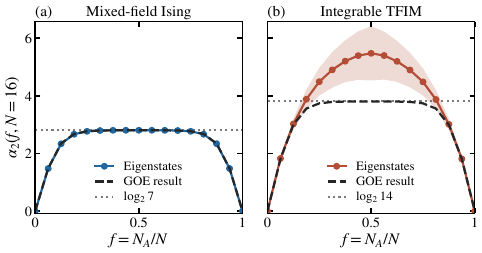}
 \caption{Stabilizer correlation versus subsystem fraction at $N=16$: (a) mixed-field Ising chain; (b) even-parity TFIM. Points average $100$ eigenstates. Shading indicate the standard deviations across eigenstates. All Pauli sums are evaluated exactly. Dashed curves show the finite-
size random-matrix reference. Dotted lines mark $\log_2 7$ and $\log_2 14$.}
 \label{fig:partition}
\end{figure}

\textit{Numerical verification.---}
We now verify these analytical results numerically.
We use the Ising chain with transverse and longitudinal fields,
\begin{equation}
 \hat H=\sum_{i=1}^{N-1}\hat Z_i\hat Z_{i+1}+\sum_{i=1}^Ng_i\hat X_i+h\sum_{i=1}^N\hat Z_i,
 \label{eq:model}
\end{equation}
with open boundaries and site Pauli matrices $\hat X_i,\hat Y_i,\hat Z_i$. The Ising coupling sets the energy unit. We choose $g_1=-1$, $g_{i>1}=(5+\sqrt5)/8$ to break reflection symmetry, and $h=(1+\sqrt5)/4$, giving a nonintegrable Hamiltonian with no additional unitary symmetry. At $h=0$, we obtain the integrable transverse-field Ising model (TFIM), for which we select the even spin-inversion sector $\hat U|\psi\rangle=|\psi\rangle$. We average the $\min(100,D-1)$ eigenstates near the middle of the spectrum, where $D$ is the sector dimension. Region $A$ contains the first $N_A$ sites.
To contrast the chaotic eigenstates with strongly disordered states, we also consider
$\hat H_W=\hat{H}+\sum_{i=1}^N w_i\hat Z_i,$
where $\hat H$ is the open mixed-field chain in Eq.~\eqref{eq:model}. The random fields $w_i$ are drawn independently from $[-6,6]$. We average $100$ independent disorder realizations, using up to $50$ eigenstates in each realization.
We evaluate the complete Pauli sums for the full state and both subsystems at each bipartition using a Walsh--Hadamard transform~\cite{xor,hadamag}.

\begin{figure}[tbp]
 \includegraphics[width=\columnwidth]{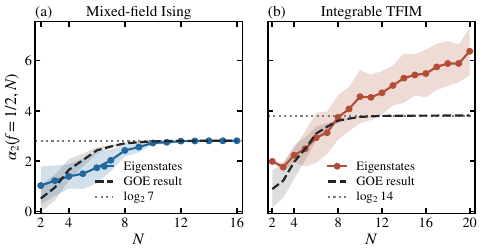}
 \caption{Size dependence of $\alpha_2(f,N)$ at $N_A=\lfloor N/2\rfloor$. (a) Mixed-field Ising chain, up to $N=16$. (b) Even-parity TFIM, up to $N=20$. Points show eigenstate averages and shading indicate the standard
deviations across eigenstates. All Pauli sums are evaluated exactly. Dashed curves show the finite-size random-matrix reference, and dotted lines mark $\log_2 7$ and $\log_2 14$.}
 \label{fig:size}
\end{figure}

\textit{mSRE and mutual information.—}
Figure~\ref{fig:entropies} shows $I_2(f,14)$ and $\mathcal L_2(f,14)$ for chaotic and integrable eigenstates. At first glance, both quantities exhibit qualitatively similar dependence on $f$ in chaotic and integrable systems, with the most apparent difference being the magnitude of eigenstate-to-eigenstate fluctuations, as indicated by the blue and orange bands. However, the distinction between the two systems becomes evident in their difference, $I_2-\mathcal L_2$. In the chaotic chain, this difference closely follows the GOE prediction, $\log_2 7$, over the range $5/14\le f\le9/14$. In contrast, for the integrable chain, the difference deviates from the random-matrix prediction and does not remain constant. This highlights the stabilizer correlation as a key quantity for distinguishing chaotic from integrable systems. In the following, we focus on the stabilizer correlation and investigate its behavior in greater detail.

\textit{Stabilizer correlation of eigenstates.---}
Figure~\ref{fig:partition}(a) shows $\alpha_2(f,N=16)$ for the mixed-field chain. It is nearly constant for $6/16\le f\le10/16$, approching random matrix reference $\log_2 7$. At $f=1/2$, its value is approximately $2.80$. As $f$ approaches $0$ or $1$, one subsystem becomes small
and $\alpha_2(f,16)$ decreases below the central plateau.
The full random-state expression, Eq.~\eqref{eq:finite}, describes this decrease, with a maximum difference of $0.00989$ from the numerical means over all $f$. Thus the decrease away from $f=1/2$ is consistent with the analytical results. As $N$ increases, we expect $\alpha_2(f,N)$ to approach $\log_2 7$ at any fixed $0<f<1$, as illustrated by the random-state curves in Fig.~\ref{fig:random}.
Figure~\ref{fig:size}(a) shows the dependence on $N$ at $N_A=\lfloor N/2\rfloor$. The stabilizer correlation increases with system size and
saturates the random-state prediction, reaching
$\alpha_2(1/2,16)\approx2.80$. The narrowing shaded band shows the suppression of eigenstate-to-eigenstate fluctuations with increasing $N$. The random-matrix prediction for the standard deviation across eigenstates is $\sigma_{\text{GOE}}\approx 7.5/D$, as derived in the SM~\cite{supplement}. At $N=16$, the numerical standard deviation is $4.48\times10^{-4}$, which is of the same order as the random-matrix prediction. At these sizes the band is narrower than the symbols. Both the weak $f$ dependence when the two subsystems are large and the approach with increasing $N$ support the constant relation in chaotic systems obtained above.

For the integrable TFIM, $\alpha_2(f,16)$ increases as $f$ approaches $1/2$ [Fig.~\ref{fig:partition}(b)]. Near equal bipartition it exceeds the limiting random-state value marked by the dotted line. The broader shaded band shows larger eigenstate-to-eigenstate fluctuations than in the chaotic chain. At $f=1/2$, the standard deviation across eigenstates is $0.941$ for the TFIM, much larger than the random-matrix prediction. In Fig.~\ref{fig:size}(b), the stabilizer correlation in the TFIM continues
to increase over the system sizes studied, retaining
a pronounced size dependence in contrast to the chaotic chain. Thus the TFIM eigenstates are not typical of the random-state ensemble.
Figure~\ref{fig:disorder} shows $\alpha_2(f,N)$ in the strongly disordered chain. The stabilizer correlation remains much smaller than the chaotic value $\log_2 7$ across the subsystem fractions. This behavior can be understood from the product-state limit. As a state approaches a product state, $\hat\rho_{AB}\approx\hat\rho_A\otimes\hat\rho_B$, the full-system Pauli sum in Eq.~\eqref{eq:identity} factorizes into the two subsystem sums, giving $\alpha_2\approx0$. At $N=12$ and $f=1/2$, we also find
$I_2=0.120$ and $\mathcal L_2=0.0266$.
These small values indicate weak correlations across
the bipartition and are consistent with the localized
regime expected at strong disorder \cite{serbyn}.

\begin{figure}[tbp]
 \includegraphics[width=\columnwidth]{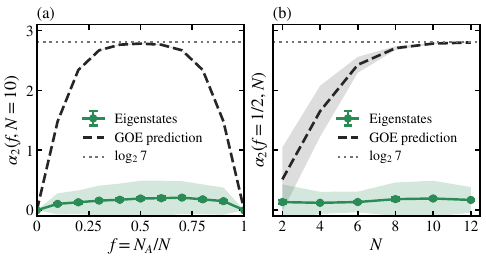}
 \caption{Stabilizer correlation in the strongly disordered Ising chain with $w_i\in[-6,6]$. (a) $\alpha_2(f,10)$ as a function of $f$. (b) $\alpha_2(1/2,N)$ as a function of system size $N$. Shading indicate the standard deviations across eigenstates; error bars show the standard error of the disorder mean. Dashed curves show the GOE random-state calculation, and dotted lines mark $\log_2 7$.}
 \label{fig:disorder}
\end{figure}

These results provide a clear physical distinction between chaotic and integrable eigenstates. When both subsystems are sufficiently large, changes in mutual information in chaotic chains are accompanied by nearly equal changes in nonstabilizerness. Consequently, both quantities exhibit volume-law scaling with the same slope as a function of system size. In contrast, for free-fermion integrable eigenstates, nonstabilizerness grows more slowly than mutual information with increasing system size.

For  interacting integrable XXZ chain, our numerical calculation also indicate an constant of the stabilizer correlation. However, its value is non-universal, and is clearly above the random-state prediction with much larger fluctuations, as shown in the 
SM~\cite{supplement}.

\textit{Summary and outlook.---}
The random-state calculation yields a universal constant difference between mutual information and mSRE in the large-subsystem limit, with values given in Table~\ref{tab:symmetry}. Exact mixed-field Ising calculations support these predictions, whereas integrable and strongly disordered eigenstates exhibit distinct profiles and size dependence. Thus, in the chaotic system, correlation growth is completely reflected in the mSRE while the stabilizer correlation approaches a constant.


\begin{acknowledgments}
W.Z. acknowledges support from the NSF of China (Grants
No. GG2030007011 and No. GG2030040453) and Innovation
Program for Quantum Science and Technology (Grant No.
2021ZD0302004). 
\end{acknowledgments}

\makeatletter
\close@column
\makeatother
\clearpage

\makeatletter
\frontmatter@init
\let\author\SupplementAuthor
\let\affiliation\SupplementAffiliation
\let\email\SupplementEmail
\let\maketitle\SupplementMakeTitle
\let\original@label\label
\def\supplement@firstpage{FirstPage}
\def\supplement@lastbibitem{LastBibItem}
\def\label#1{%
  \def\supplement@label{#1}%
  \ifx\supplement@label\supplement@firstpage
    \original@label{sm:FirstPage}%
  \else
    \ifx\supplement@label\supplement@lastbibitem
      \original@label{sm:LastBibItem}%
    \else
      \original@label{#1}%
    \fi
  \fi
}
\def\NAT@bibsetnum#1{%
  \setlength{\topsep}{\z@}%
  \NATx@bibsetnum{\ref{sm:LastBibItem}}%
}
\let\@bibsetup\NAT@bibsetnum
\makeatother
\ExplSyntaxOn
\cs_new:Npn \SupplementDestName #1
  {
    \str_if_eq:eeTF { \exp_args:Ne \str_range:nnn {#1}{1}{12} } {frontmatter.}
      {sm.#1} {#1}
  }
\ExplSyntaxOff
\let\SupplementDestFilter\HyperDestNameFilter
\renewcommand{\maketitle}{%
  \let\HyperDestNameFilter\SupplementDestName
  \SupplementMakeTitle
  \let\HyperDestNameFilter\SupplementDestFilter
}
\setcounter{page}{1}
\setcounter{equation}{0}
\setcounter{figure}{0}
\setcounter{table}{0}
\setcounter{section}{0}
\setcounter{subsection}{0}
\raggedbottom
\setlength{\abovecaptionskip}{4pt}
\renewcommand{\theequation}{S\arabic{equation}}
\renewcommand{\thefigure}{S\arabic{figure}}
\renewcommand{\thetable}{S\arabic{table}}
\renewcommand{\thesection}{S.\Roman{section}}
\renewcommand{\thesubsection}{\thesection.\Alph{subsection}}
\setcounter{secnumdepth}{3}

\title{Supplemental Material for: Universal stabilizer correlations in chaotic eigenstates}
\author{Zhi-Han Liu}
\affiliation{Hefei National Research Center for Physical Sciences at the Microscale and School of Physical Sciences, University of Science and Technology of China, Hefei 230026, China}
\author{Hao-Yue Qi}
\email[Corresponding author: ]{qhaoyue@mail.ustc.edu.cn}
\affiliation{Hefei National Research Center for Physical Sciences at the Microscale and School of Physical Sciences, University of Science and Technology of China, Hefei 230026, China}
\author{Wei Zheng}
\email[Corresponding author: ]{zw8796@ustc.edu.cn}
\affiliation{Hefei National Research Center for Physical Sciences at the Microscale and School of Physical Sciences, University of Science and Technology of China, Hefei 230026, China}
\affiliation{CAS Center for Excellence in Quantum Information and Quantum Physics, University of Science and Technology of China, Hefei 230026, China}
\affiliation{Hefei National Laboratory, University of Science and Technology of China, Hefei 230088, China}
\maketitle

\section{Notation and random-state reference}
\label{sec:definitions}
Consider a pure $N$-qubit state $\hat\rho_{AB}=|\psi\rangle\langle\psi|$ with $N_A+N_B=N$ and $f=N_A/N$. For $\Omega=A,B,AB$, let $\hat\rho_\Omega$ denote its reduced state, $N_{AB}=N$, and $d_\Omega=2^{N_\Omega}$, so $d=d_Ad_B=2^N$. The phase-free Pauli strings are $\hat P_\Omega\in\Pauli_{N_\Omega}=\{\hat I,\hat X,\hat Y,\hat Z\}^{\otimes N_\Omega}$. Define
\begin{equation}
 F_\Omega=\sum_{\hat P_\Omega\in\Pauli_{N_\Omega}}x_{\hat P_\Omega}^4,
 \qquad x_{\hat P_\Omega}=\Tr(\hat\rho_\Omega\hat P_\Omega),
 \label{eq:sectorF}
\end{equation}
so that the stabilizer correlation is
\begin{equation}
 \alpha_2=\log_2\frac{F_{AB}}{F_AF_B}.
 \label{eq:alpha}
\end{equation}

For a symmetry sector with projector $\hat\Pi$ and dimension $D$, we consider real and complex random states corresponding to the Gaussian orthogonal and unitary ensembles (GOE and GUE), respectively. With $\EE$ denoting the ensemble average, define the random-state reference
\begin{equation}
 \alpha_2^{\rm RMT}=\log_2\frac{\EE F_{AB}}{(\EE F_A)(\EE F_B)}.
 \label{sm:eq:momentreference}
\end{equation}
At fixed $0<f<1$, the concentration analysis in Sec.~\ref{app:concentration} establishes that a typical random state's $\alpha_2$ and the ensemble mean $\EE\alpha_2$ approach the same large-system limit as $\alpha_2^{\rm RMT}$.

\section{Random-state formulas and concentration}
\label{app:random}
We derive the fourth moments of Pauli expectation values from normalized Gaussian amplitudes and sum them over Pauli strings to obtain the ensemble-averaged Pauli sums. The final subsection bounds fluctuations of these sums to obtain the stabilizer correlation of a typical random state.

\subsection{Exact moments of random states}
\label{app:exact}
In a general orthonormal basis $\{|i\rangle\}_{i=1}^D$ of a
$D$-dimensional symmetry sector, a random state can be written as
$|\psi\rangle=\sum_{i=1}^D z_i|i\rangle$.
Here, $z_i=g_i/\sqrt r$, where the $g_i$ are independent
Gaussian random variables with zero mean and unit variance
($\EE|g_i|^2=1$), and $r=\sum_i|g_i|^2$.
The $g_i$ are real for the GOE and complex for the GUE.
To treat the GOE and GUE cases together, we define $c=3-\beta$,
where the Dyson index is $\beta=1$ for the GOE and $\beta=2$ for the GUE. Thus $c=2$ for real states and $c=1$ for complex states.
Let $\hat A$ denote the restricted Pauli operator: $\hat A=\operatorname{Re}(\hat\Pi\hat P\hat\Pi)$ in a real orthonormal sector basis for the GOE, and $\hat A=\hat\Pi\hat P\hat\Pi$ for the GUE. For a real projector, odd-$\hat Y$ strings have $\hat A=0$. The eigenbasis of $\hat A$ can be chosen real in the GOE case. In this eigenbasis,
\begin{equation}
 \begin{aligned}
 x_{\hat P}&=\sum_{i=1}^D\lambda_i|z_i|^2=\frac{U}{r},\\
 U&=\sum_i\lambda_i|g_i|^2,
 \end{aligned}
 \label{eq:pauliamplitudes}
\end{equation}
where $\lambda_i$ are the eigenvalues of $\hat A$. Since $r$ is independent of $z_i$, we have $\EE U^4=(\EE r^4)(\EE x_{\hat P}^4)$. We can therefore obtain $\EE x_{\hat P}^4=\EE U^4/(\EE r^4)$.
We obtain $\EE U^4$ from the moment-generating function $\EE e^{sU}$. Independence of the $g_i$ gives 
\begin{equation}\label{prod}
    \EE e^{sU}=\prod_i(1-cs\lambda_i)^{-1/c}.
\end{equation}
Setting all $\lambda_i=1$ gives
\begin{equation*}
 \EE r^4=D(D+c)(D+2c)(D+3c).
\end{equation*}

For the full-space and spin-inversion calculations, each
operator $\hat A$ whose expectation value varies among states
in the ensemble has $D/2$ eigenvalues $+1$ and $D/2$
eigenvalues $-1$. Pairing these factors in the product gives
\begin{equation*}
 \begin{aligned}
 \EE e^{sU}&=(1-c^2s^2)^{-D/(2c)},\\
 \EE U^4&=3c^2D(D+2c).
 \end{aligned}
\end{equation*}
Dividing by $\EE r^4$ therefore gives
\begin{equation}
 \EE x_{\hat P}^4=\frac{3c^2}{(D+c)(D+3c)}.
 \label{eq:involutionfourth}
\end{equation}
Setting $c=2$ for GOE or $1$ for GUE gives $12/[(D+2)(D+6)]$ or $3/[(D+1)(D+3)]$, respectively. 

Pauli strings
restricted to the translation, reflection, or magnetization symmetry sectors can have other eigenvalues. To calculate $\EE x_{\hat P}^4$ in these symmetry sectors, we use the same product formula, Eq.~\eqref{prod},
for $\EE e^{sU}$ as above, but now keep the eigenvalues
arbitrary. Expanding the product and collecting the $s^4$
term gives $\EE U^4/4!$. Dividing $\EE U^4$ by
$\EE r^4=D(D+c)(D+2c)(D+3c)$ then yields
\begin{equation}
 \begin{aligned}
 \EE x_{\hat P}^4
 ={}&
 \frac{1}{D(D+c)(D+2c)(D+3c)}\\
 &\times\Bigl[
 (\Tr\hat A)^4
 +6c(\Tr\hat A)^2\Tr(\hat A^2)\\
 &\qquad
 +3c^2[\Tr(\hat A^2)]^2
 +8c^2(\Tr\hat A)\Tr(\hat A^3)\\
 &\qquad
 +6c^3\Tr(\hat A^4)
 \Bigr].
 \end{aligned}
 \label{eq:projectedfourth}
\end{equation}
Here, we have written the eigenvalue
sums $\sum_i\lambda_i^j=\Tr(\hat A^j)$ as traces taken over the sector.

\subsection{Counting the Pauli strings in full space}
\label{app:finite}
\label{app:purity}
Every nonidentity even-$\hat Y$ Pauli string is symmetric,
traceless, and squares to the identity. Embedded in the full
Hilbert space it has $d/2$ eigenvalues of each sign. Equation~\eqref{eq:involutionfourth} therefore gives
\begin{equation}
 \EE x_{\hat P_\Omega}^4=\frac{12}{(d+2)(d+6)}.
 \label{eq:fullfourth}
\end{equation}
An odd-$\hat Y$ string is antisymmetric under transposition
and has zero expectation in a real state.
For real random states in the full Hilbert space, only
Pauli strings with even $Y$ parity contribute to $\EE F_\Omega$.
There are three non-$Y$ choices and one $Y$ choice at each
site, so the number of even-$Y$ strings, including the
identity, is
\begin{equation*}
 \frac{(3+1)^{N_\Omega}+(3-1)^{N_\Omega}}{2}
 =
 \frac{d_\Omega^2+d_\Omega}{2}.
\end{equation*}
The identity contributes one, while each remaining string
contributes the fourth moment in
Eq.~\eqref{eq:involutionfourth}. For real states in the full Hilbert space, $D=d$ and $c=2$, giving
\begin{equation}
 \begin{aligned}
 \EE F_\Omega
 &=
 1+\left(\frac{d_\Omega^2+d_\Omega}{2}-1\right)
 \frac{12}{(d+2)(d+6)}\\
 &=
 1+\frac{6(d_\Omega^2+d_\Omega-2)}{(d+2)(d+6)}.
 \end{aligned}
\label{eq:mainfullmeans}
\end{equation}

For complex random states, all $d_\Omega^2-1$ nonidentity strings
have the same fourth moment, $3/[(d+1)(d+3)]$.
Adding their contributions to the identity term gives
\begin{equation*}
 \EE F_\Omega
 =
 1+\frac{3(d_\Omega^2-1)}{(d+1)(d+3)}.
\end{equation*}
The full-system nonidentity sum tends to $3$ and the
identity adds $1$, while both subsystem sums tend to $1$.
The limiting stabilizer correlation for complex random
states is therefore $\log_2 4$.

With $d_{AB}=d$, substitution into Eq.~\eqref{sm:eq:momentreference} yields
\begin{equation}
 \alpha_2^{\rm RMT}(f,N)=\log_2
 \frac{7d/(d+6)}
 {\displaystyle\prod_{\Omega=A,B}\left[1+\frac{6(d_\Omega^2+d_\Omega-2)}{(d+2)(d+6)}\right]}.
 \label{sm:eq:finite}
\end{equation}
At fixed $0<f<1$, both subsystems grow with $N$, giving $\EE F_A,\EE F_B\to1$, whereas $\EE F_{AB}\to7$. Hence we have
\begin{equation}
 \alpha_2^{\rm RMT}(f,N)=\log_2 7-O\!\left(2^{-N}+2^{-2N_A}+2^{-2N_B}\right).
 \label{eq:gaussianoneside}
\end{equation}

\subsection{Counting the Pauli strings in spin-inversion sectors}
\label{app:spin}
We consider random states in a fixed spin-inversion sector of a Hamiltonian satisfying
\begin{equation}
 \begin{aligned}
 \hat U&=\prod_{i=1}^N\hat X_i,
 \qquad [\hat H,\hat U]=0,\\
 \hat\Pi_s&=\frac{\hat I+s\hat U}{2},
 \qquad s=\pm1.
 \end{aligned}
 \label{eq:spinoperator}
\end{equation}
The sector is defined by $\hat U|\psi\rangle=s|\psi\rangle$ and has dimension $D=d/2$. Throughout this subsection, both subsystems are nonempty, $N_A,N_B\geq1$.

The relation $\hat U\hat P\hat U=(-1)^{\#Y+\#Z}\hat P$ shows that a Pauli string with an odd total number of $\hat Y$ and $\hat Z$ factors anticommutes with $\hat U$. It maps the state to the opposite sector and therefore has zero expectation. For real states, strings with an odd number of $\hat Y$ factors also have zero expectation, so the numbers of $\hat Y$ and $\hat Z$ factors must each be even.

For the full system, $\hat I$ and $\hat U$ have fixed expectations $1$ and $s$, respectively, and each contributes one to $F_{AB}$. A proper subsystem $\Omega=A,B$ contains only the identity as a fixed nonzero contribution. Every other allowed string, embedded as $\hat P_\Omega\otimes\hat I_{\bar\Omega}$ for a subsystem observable, is traceless within the sector and squares to the identity there. Its restricted spectrum thus contains $D/2$ eigenvalues of each sign. Equation~\eqref{eq:involutionfourth} applies with the same sector dimension $D=d/2$ to both full-system and subsystem strings.

For real states, the two parity conditions allow $(d_\Omega^2+2d_\Omega)/4$ strings, including the fixed contributions. Each nonfixed allowed string has fourth moment $48/[(d+4)(d+12)]$. Subtracting the two fixed strings for the full system, or the single identity for a proper subsystem, gives
\begin{equation}
 \begin{aligned}
 \EE F_{AB}
 &=2+\left(\frac{d^2+2d}{4}-2\right)
 \frac{48}{(d+4)(d+12)}\\
 &=\frac{14d}{d+12},\\
 \EE F_\Omega
 &=1+\left(\frac{d_\Omega^2+2d_\Omega}{4}-1\right)
 \frac{48}{(d+4)(d+12)}\\
 &=1+\frac{12(d_\Omega^2+2d_\Omega-4)}{(d+4)(d+12)}, \qquad \Omega=A,B.
 \end{aligned}
 \label{eq:spinfull}
\end{equation}

For complex states, only the total $(Y+Z)$ parity is restricted, leaving $d_\Omega^2/2$ allowed strings. The fourth moment of each nonfixed string is $12/[(d+2)(d+6)]$. The corresponding sums are
\begin{equation}
 \begin{aligned}
 \EE F_{AB}
 &=2+\left(\frac{d^2}{2}-2\right)
 \frac{12}{(d+2)(d+6)}\\
 &=\frac{8d}{d+6},\\
 \EE F_\Omega
 &=1+\left(\frac{d_\Omega^2}{2}-1\right)
 \frac{12}{(d+2)(d+6)}\\
 &=1+\frac{6(d_\Omega^2-2)}{(d+2)(d+6)}, \qquad \Omega=A,B.
 \end{aligned}
\end{equation}

Substituting the real-state sums into Eq.~\eqref{sm:eq:momentreference} gives the exact finite-size GOE reference,
\begin{equation}
 \alpha_2^{\rm RMT}(f,N)
 =\log_2
 \frac{14d/(d+12)}
 {\displaystyle\prod_{\Omega=A,B}
 \left[1+\frac{12(d_\Omega^2+2d_\Omega-4)}
 {(d+4)(d+12)}\right]}.
 \label{eq:spinfinite}
\end{equation}
The complex-state reference follows from the same logarithmic ratio using the complex-state sums above. The GOE expression gives the TFIM random-state reference curves.

At fixed $0<f<1$, both subsystems grow with $N$. The subsystem means approach one, while the full-system mean approaches $14$ for real states and $8$ for complex states. Consequently, the GOE reference obeys
\begin{equation}
 \alpha_2^{\rm RMT}(f,N)=\log_2 14
 -O\!\left(2^{-N}+2^{-2N_A}+2^{-2N_B}\right).
 \label{eq:spinconstant}
\end{equation}
The finite-size corrections are exponentially small, while the GUE reference approaches $\log_2 8$. Section~\ref{app:concentration} shows that fluctuations of the Pauli sums around their ensemble averages vanish in the large-system limit. Thus a typical random state's $\alpha_2$ and the mean $\EE\alpha_2$ approach these same constants.

\subsection{Translation symmetry}
\label{app:translation}
We calculate the Pauli sums for random states with fixed
momentum. The translation operator acts on spin product states as
$\hat T|s_1\cdots s_N\rangle=|s_Ns_1\cdots s_{N-1}\rangle$.
The sector with quantum number $e^{i2\pi k/N}$ has projector
\begin{equation}
 \hat\Pi_k=\frac1N\sum_{r=0}^{N-1}
 e^{-i2\pi kr/N}\hat T^r,
 \qquad k=0,\ldots,N-1.
 \label{eq:momentumprojector}
\end{equation}
Real random states are possible only at $k=0$ and, for even
$N$, $k=N/2$; complex random states can occupy any sector.
A shift by $r$ has $\gcd(N,r)$ cycles. Each cycle admits
two invariant spin configurations, so
$\Tr\hat T^r=2^{\gcd(N,r)}$. Inserting a Pauli string
replaces each factor of two by the trace of a product of
single-site Pauli matrices, whose magnitude is at most two.
Since $\gcd(N,r)\leq N/2$ for $r\neq0$, we obtain
\begin{equation}
 \begin{aligned}
 D=\Tr\hat\Pi_k
 &=\frac1N\sum_{r=0}^{N-1}
 e^{-i2\pi kr/N}2^{\gcd(N,r)}\\
 &=\frac dN\left[1+O\!\left(\frac N{\sqrt d}\right)\right],\\
 |\Tr(\hat\Pi_k\hat P)|&\leq\sqrt d,
 \qquad \hat P\neq\hat I.
 \end{aligned}
 \label{eq:translationdimension}
\end{equation}

To calculate $\EE F_{AB}$, we sum the exact fourth moment
in Eq.~\eqref{eq:projectedfourth} over all Pauli strings.
Rather than evaluate each projected
operator separately, we use Pauli orthogonality to sum
the trace terms in the numerator:
\begin{equation}
 \begin{aligned}
 \sum_{\hat P}(\Tr\hat A)^2&=dD,\\
 \sum_{\hat P}\Tr(\hat A^2)&\leq dD^2,\\
 \sum_{\hat P}[\Tr(\hat A^2)]^2
 &=\frac1c\sum_{\hat P}(\Tr\hat A)^4+O(dD^2).
 \end{aligned}
 \label{eq:paulisumidentities}
\end{equation}
All sums run over $\Pauli_N$, and traces of $\hat A$ are
taken over the sector; for real states, odd-$\hat Y$ strings contribute zero by definition. The first identity follows from
the Pauli expansion of the projector; the second follows
from $\sum_{\hat P}\hat P\hat\Pi_k\hat P=dD\hat I$.
For the third relation, define
$a_{\hat P}=\Tr(\hat\Pi_k\hat P)$ and
$b_{\hat P}=\Tr[(\hat\Pi_k\hat P\hat\Pi_k)^2]$.
Pauli orthogonality gives $\sum_{\hat P}b_{\hat P}^2=\sum_{\hat P}a_{\hat P}^4$.
For a real projector, keeping only the real symmetric parts instead gives
\begin{equation*}
 \sum_{\hat P}[\Tr(\hat A^2)]^2
 =\frac12\sum_{\hat P}a_{\hat P}^4
 +\frac12\sum_{\hat P}a_{\hat P}^2b_{\hat P}.
\end{equation*}
The last term is bounded by $dD^2/2$, since
$0\leq b_{\hat P}\leq D$ and $\sum_{\hat P}a_{\hat P}^2=dD$.
This proves the third relation for $c=1$ and $c=2$.
It combines the two numerator terms
$(\Tr\hat A)^4$ and $3c^2[\Tr(\hat A^2)]^2$ into
$(1+3c)\sum_{\hat P}(\Tr\hat A)^4$, up to $O(dD^2)$.
The first two identities, together with $\|\hat A\|\leq1$
and Cauchy--Schwarz, bound the sum of the other three
terms by $O(dD^2)$. Thus
\begin{equation*}
 \EE F_{AB}
 =
 \frac{
 (1+3c)\sum_{\hat P}(\Tr\hat A)^4+O(dD^2)
 }{D(D+c)(D+2c)(D+3c)}.
\end{equation*}

For a subsystem $\Omega$, we instead sum only over
$\hat P=\hat P_\Omega\otimes\hat I_{\bar\Omega}$.
The complete-basis identity for
$\sum_{\hat P}[\Tr(\hat A^2)]^2$ no longer applies.
Since there are $d_\Omega^2$ subsystem Pauli strings and
$\Tr(\hat A^2)\leq D$, we use the bound
\begin{equation*}
 \sum_{\hat P_\Omega}[\Tr(\hat A^2)]^2\leq d_\Omega^2D^2.
\end{equation*}
The other three terms obey the same total bounds as in
the full-system calculation. Using
$\EE x_{\hat P}=\Tr\hat A/D$, we therefore obtain
\begin{equation}
 \begin{aligned}
 \EE F_{AB}
 &=(1+3c)\sum_{\hat P\in\Pauli_N}
 (\EE x_{\hat P})^4+O\!\left(\frac d{D^2}\right),\\
 \EE F_\Omega
 &=\sum_{\hat P_\Omega\in\Pauli_{N_\Omega}}
 (\EE x_{\hat P_\Omega})^4
 +O\!\left(\frac d{D^2}+\frac{d_\Omega^2}{D^2}\right).
 \end{aligned}
 \label{eq:paulisumsfrommeans}
\end{equation}
Replacing the normalization denominator by $D^4$
introduces an error of at most $O(d/D^2)$, since
$\sum_{\hat P}(\Tr\hat A)^4\leq dD^3$. Since the derivation uses only Pauli orthogonality and
the properties of the sector projector, rather than
translation symmetry, Eq.~\eqref{eq:paulisumsfrommeans}
also applies to other symmetry sectors.

It remains to evaluate the sums of fourth powers of
$\EE x_{\hat P}$ for the momentum sector.
The identity contributes one. For every nonidentity
string, Eq.~\eqref{eq:translationdimension} gives
$|\Tr\hat A|\leq\sqrt d$, so
\begin{equation*}
 \begin{aligned}
 \sum_{\hat P\neq\hat I}(\EE x_{\hat P})^4
 &=\frac1{D^4}
 \sum_{\hat P\neq\hat I}(\Tr\hat A)^4\\
 &\leq\frac d{D^4}
 \sum_{\hat P}(\Tr\hat A)^2
 =\frac{d^2}{D^3}.
 \end{aligned}
\end{equation*}
Substituting
$D\sim d/N$ and $d=d_\Omega d_{\bar\Omega}$ into
Eq.~\eqref{eq:paulisumsfrommeans} gives
\begin{equation}
 \begin{aligned}
 \EE F_{AB}
 &=1+3c+O\!\left(\frac{N^3}{d}\right),\\
 \EE F_\Omega
 &=1+O\!\left(
 \frac{N^3}{d}+\frac{N^2}{d_{\bar\Omega}^{\,2}}
 \right),
 \qquad \Omega=A,B.
 \end{aligned}
 \label{eq:translationmeans}
\end{equation}
At fixed $0<f<1$, all these errors vanish:
the full-system mean approaches $1+3c$, while both subsystem
means approach one. Equation~\eqref{sm:eq:momentreference}
therefore gives $\alpha_2^{\rm RMT}\to\log_2(1+3c)$,
namely $\log_2 7$ for real states and $\log_2 4$ for
complex states.

\subsection{Reflection symmetry}
\label{app:reflection}

Reflection acts as
$\hat R|s_1\cdots s_N\rangle=|s_N\cdots s_1\rangle$.
The sector projector and dimension are
\begin{equation}
 \hat\Pi_s=\frac{\hat I+s\hat R}{2},
 \qquad
 D=\frac{d+s2^{\lceil N/2\rceil}}2\sim\frac d2,
 \qquad s=\pm1.
 \label{eq:reflectionprojector}
\end{equation}
Each mirror pair contributes at most two to
$\Tr(\hat R\hat P)$, as does the central site when $N$
is odd. For a nonidentity string,
\begin{equation*}
 |\Tr(\hat\Pi_s\hat P)|
 =\frac12|\Tr(\hat R\hat P)|
 \leq2^{\lceil N/2\rceil-1}=O(\sqrt d).
\end{equation*}
Together with $\sum_{\hat P}(\Tr\hat A)^2=dD$, this
bounds the nonidentity sum of $(\EE x_{\hat P})^4$ by
$O(d^{-1})$. Equation~\eqref{eq:paulisumsfrommeans} applies to a general
sector projector. We obtain
\begin{equation}
 \begin{aligned}
 \EE F_{AB}&=1+3c+O(d^{-1}),\\
 \EE F_\Omega&=1+O\!\left(d^{-1}+d_{\bar\Omega}^{-2}\right),
 \qquad \Omega=A,B.
 \end{aligned}
 \label{eq:reflectionmeans}
\end{equation}
For either parity and fixed $0<f<1$, the limiting
$\alpha_2^{\rm RMT}$ is again $\log_2 7$ for real states
and $\log_2 4$ for complex states.

\subsection{Half-filled magnetization}
\label{app:magnetization}

The effect of a fixed $U(1)$ charge on full-system SRE statistics has also been studied for constrained Haar random states \cite{sm:iannottiu1}. Here we fix the number of up spins to $N/2$, with even $N$.
The sector dimension is
\begin{equation}
 D=\binom{N}{N/2}
 =d\sqrt{\frac{2}{\pi N}}\,[1+O(N^{-1})].
 \label{eq:halfsector}
\end{equation}
Equation~\eqref{eq:paulisumsfrommeans} reduces the
calculation to ensemble means.
A string containing $\hat X$ or $\hat Y$ has zero diagonal
and hence zero mean. For a diagonal string with $\ell$
factors of $\hat Z$, counting spin configurations gives
\begin{equation}
 \EE x_{\hat P}=
 \begin{cases}
 \displaystyle(-1)^j\frac{\binom{N/2}{j}}{\binom{N}{2j}},
 &\ell=2j,\\[2mm]
 0,&\ell\text{ odd}.
 \end{cases}
 \label{eq:halfdiagonalmean}
\end{equation}

There are $\binom N\ell$ such full-system strings and
$\binom{N_\Omega}\ell$ in region $\Omega$.
The strings with $\ell=0,N$ have fixed expectations of
magnitude one. The leading corrections come from
$\ell=2,N-2$, with mean magnitude $1/(N-1)$.
A subsystem contains only the $\ell=0$ fixed term
and the $\ell=2$ leading correction. At fixed $0<f<1$,
summing the remaining weights gives $O(N^{-4})$, so
\begin{equation}
 \begin{aligned}
 \sum_{\hat P\in\Pauli_N}(\EE x_{\hat P})^4
 &=2+\frac{2\binom N2}{(N-1)^4}+O(N^{-4}),\\
 \sum_{\hat P_\Omega\in\Pauli_{N_\Omega}}(\EE x_{\hat P_\Omega})^4
 &=1+\frac{\binom{N_\Omega}2}{(N-1)^4}+O(N^{-4}).
 \end{aligned}
 \label{eq:halfdiagonalsums}
\end{equation}
Substituting into Eq.~\eqref{eq:paulisumsfrommeans} yields
\begin{equation}
 \begin{aligned}
 \EE F_{AB}
 &=(1+3c)\left[2+\frac{2\binom N2}{(N-1)^4}\right]
 +O\!\left(N^{-4}+\frac Nd\right),\\
 \EE F_\Omega
 &=1+\frac{\binom{N_\Omega}2}{(N-1)^4}
 +O\!\left(N^{-4}+\frac Nd+\frac N{d_{\bar\Omega}^{\,2}}\right).
 \end{aligned}
 \label{eq:halfmeans}
\end{equation}
Taking their logarithmic ratio gives
\begin{equation}
 \alpha_2^{\rm RMT}(f,N)
 =\log_2[2(1+3c)]
 +\frac{f(1-f)}{N^2\ln2}+O(N^{-3}).
 \label{eq:halffillingcorrection}
\end{equation}
The limits are $\log_2 14$ for real states and $\log_2 8$
for complex states. The $N^{-2}$ correction comes from
the nonzero two-spin mean $\EE x_{\hat Z_i\hat Z_j}
=-1/(N-1)$ for $i\neq j$.

\subsection{Concentration and the mean stabilizer correlation}
\label{app:concentration}

We now show that the mean sums determine the limiting
$\alpha_2$ of a typical random state. Strings with
state-independent expectations contribute no variance
and are omitted from the primed sums below.
On the unit sphere of state amplitudes, differentiation
and Pauli orthogonality bound the tangential gradient by
$\|\nabla_{\rm tan}F_\Omega\|^2\leq64d\sum_{\hat P_\Omega}'x_{\hat P_\Omega}^6$.
The sphere Poincar\'e inequality therefore yields
\begin{equation}
 \operatorname{Var}F_\Omega
 \leq\frac{64d}{D-1}
 \sum_{\hat P_\Omega}'\EE x_{\hat P_\Omega}^6,
 \qquad D>1,
 \quad \Omega=A,B,AB.
 \label{eq:paulisumvariance}
\end{equation}
For complex states, the unit sphere lies in $\mathbb R^{2D}$
and the denominator is $2D-1$, so the bound above also holds.

The normalized-Gaussian calculation at sixth order,
together with $\|\hat A\|\leq1$, gives
$\EE|x_{\hat P}-\EE x_{\hat P}|^6=O(D^{-3})$ uniformly
over the strings. Since there are at most $d^2$ strings,
\begin{equation*}
 \operatorname{Var}F_\Omega
 =O\!\left[
 \frac{d^3}{D^4}
 +\frac dD\sum_{\hat P_\Omega}'(\EE x_{\hat P_\Omega})^6
 \right].
\end{equation*}
The primed means vanish in the full-space and
spin-inversion ensembles. The trace bounds above and
$\sum_{\hat P}(\EE x_{\hat P})^2=d/D$ bound their sixth-power
sum by $O(N^5/d^2)$ for momentum sectors and $O(d^{-2})$
for reflection sectors. At half filling,
Eq.~\eqref{eq:halfdiagonalmean} instead gives $O(N^{-4})$,
with leading contributions from $\ell=2,N-2$.
Consequently, for $\Omega=A,B,AB$,
\begin{equation}
 \operatorname{Var}F_\Omega=
 \begin{cases}
 O(d^{-1}),&\substack{\text{full space, spin inversion,}\\
 \text{or reflection}},\\
 O(N^4/d),&\text{momentum},\\
 O(N^{-7/2}+N^2/d),&\text{half filling}.
 \end{cases}
 \label{eq:sectorconcentration}
\end{equation}

Since $F_\Omega\geq1$, the logarithm is Lipschitz with constant
$1/\ln2$. Thus
\begin{equation}
 \EE\left|\alpha_2-\alpha_2^{\rm RMT}\right|
 \leq\frac1{\ln2}
 \sum_{\Omega\in\{A,B,AB\}}\sqrt{\operatorname{Var}F_\Omega}
 \longrightarrow0.
 \label{eq:alphaconcentration}
\end{equation}
At fixed $0<f<1$, $\EE\alpha_2$ and a typical random
state's $\alpha_2$ approach the same constants as
$\alpha_2^{\rm RMT}$.

\subsection{GOE prediction for the finite-size standard deviation}
\label{app:goefluctuations}

We calculate the GOE standard deviation of $\alpha_2$. The reference is the full-space GOE ($D=d$) for the mixed-field Ising chain and the even spin-inversion GOE ($D=d/2$) for the TFIM.
For either sector, the subsystem means in Eqs.~\eqref{eq:mainfullmeans} and \eqref{eq:spinfull} can be written as
\begin{equation*}
 \mu_\Omega\equiv\EE F_\Omega
 =1+\frac{6[(D/d)d_\Omega^2+d_\Omega-2]}{(D+2)(D+6)},
 \qquad \Omega=A,B.
\end{equation*}
We obtain the Pauli-sum covariances by applying Eq.~\eqref{prod} to two sector-restricted operators and summing the connected fourth moments over ordered pairs. With these covariances, the first-order expansion of the logarithms in Eq.~\eqref{eq:alpha} gives the finite-size standard deviation
\begin{equation}
 \begin{aligned}
 &\sigma_{\rm GOE}(N,N_A)\\
 &\quad\simeq
 \frac{4}{\ln 2\sqrt{(D+4)(D+10)(D+12)(D+14)}}\\
 &\quad\times\Biggl[
 \frac{12(D-1)(D+4)}{7}
 -4D(D+4)(2-\mu_A^{-1}-\mu_B^{-1})\\
 &\qquad+D(D^2+24D+104)(2-\mu_A^{-1}-\mu_B^{-1})^2\\
 &\qquad+8D(D+13)\sum_{\Omega=A,B}
       \frac{\mu_\Omega-1}{\mu_\Omega^2}\\
 &\qquad+\frac{192(d-D)(D+13)}
 {(D+2)(D+6)\mu_A\mu_B}\\
 &\qquad-\frac{6D(D+16)}{d}
   \sum_{\Omega=A,B}
   \frac{d_\Omega^2}{\mu_\Omega^2}
   \left(\mu_\Omega-1\right)\\
 &\qquad+\frac{72(D+16)(d-D)}{d(D+2)(D+6)}
   \sum_{\Omega=A,B}\frac{d_\Omega^2}{\mu_\Omega^2}
 \Biggr]^{1/2}.
 \end{aligned}
 \label{eq:goefluctuations}
\end{equation}
The leading order of the standard deviation is $\sigma_{\rm GOE}\approx 7.5/D$.
For example, at $N=16$ and $N_A=8$, Eq.~\eqref{eq:goefluctuations} gives $1.1532\times10^{-4}$ for the full-space GOE and $2.3070\times10^{-4}$ for the even-parity GOE.

\section{Stabilizer correlation in the XXZ chain}
\label{app:xxz}

We compare the interacting integrable XXZ chain with a
chaotic extension and real random states in the same
symmetry sector. The periodic Hamiltonian is~\cite{sm:leblondxxz}
\begin{equation}
 \begin{aligned}
 \hat H_{\rm XXZ}={}&\sum_{i=1}^N\left(
 \hat S_i^x\hat S_{i+1}^x+\hat S_i^y\hat S_{i+1}^y
 +\Delta\hat S_i^z\hat S_{i+1}^z\right)\\
 &+\lambda\sum_{i=1}^N\left(
 \hat S_i^x\hat S_{i+2}^x+\hat S_i^y\hat S_{i+2}^y
 +\Delta'\hat S_i^z\hat S_{i+2}^z\right),
 \end{aligned}
 \label{eq:xxz}
\end{equation}
where $(\hat S_i^x,\hat S_i^y,\hat S_i^z)
=(\hat X_i,\hat Y_i,\hat Z_i)/2$, site indices are taken
modulo $N$, and the nearest-neighbor exchange sets the
energy unit. We use $\Delta=1.1$ and $\Delta'=0.5$.
The model is integrable at $\lambda=0$ and nonintegrable
at $\lambda=1$. Both are restricted to zero total
magnetization, momentum $k=0$, and even reflection and
spin-inversion parities.

The random-state reference is obtained by normalizing
vectors with independent real standard Gaussian
coefficients in a real orthonormal basis of the same
joint symmetry sector. We calculate $\alpha_2$ numerically for each random state and average the resulting values.
This reference incorporates all four symmetry constraints
simultaneously.
For $N=22,24$, subsystem Pauli sums are exact, while the
full-system sums combine exact partial sums with stratified
importance sampling ($1024$ production samples per state).
Error bars show approximate $95\%$ intervals for the mean,
including eigenstate-selection and Pauli-sampling uncertainties.

\begin{figure}[tbp]
 \includegraphics[width=\columnwidth]{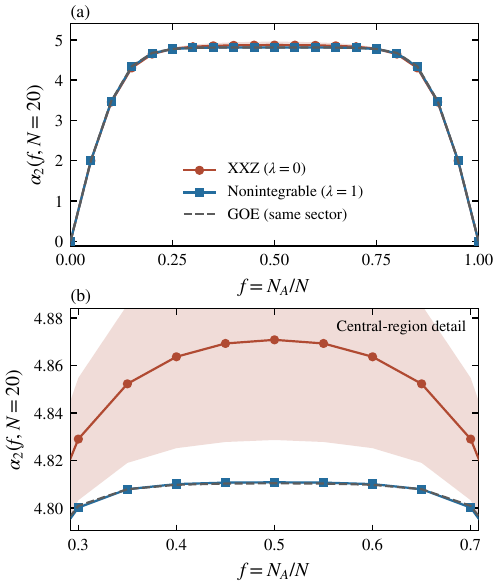}
 \caption{Stabilizer correlation at $N=20$:
 (a) dependence on subsystem fraction $f$;
 (b) magnification near $f=1/2$.
 Solid curves show averages over $192$ eigenstates near
 the spectral center for the integrable and chaotic models.
 The dashed curve averages $192$ real random states in
 the same joint symmetry sector. Shading indicate the standard deviations across eigenstates.
 All Pauli sums are evaluated exactly.}
 \label{fig:xxz_fraction}
\end{figure}
\begin{figure}[tbp]
 \includegraphics[width=\columnwidth]{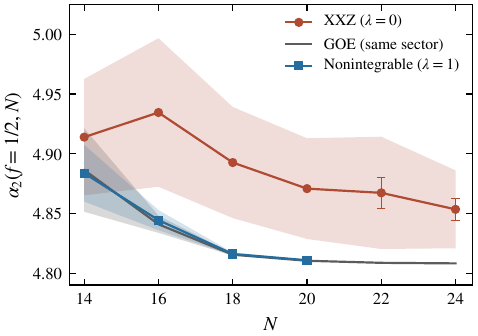}
 \caption{Half-chain stabilizer correlation versus $N$.
 Curves show means, shading indicate the standard deviations across eigenstates, and error bars indicate approximate $95\%$
 statistical intervals for Monte Carlo sampling for $N\geq 22$.}
 \label{fig:xxz_scaling}
\end{figure}

Figure~\ref{fig:xxz_fraction} shows that both models develop
a plateau near $f=1/2$, unlike the pronounced central peak
in the free-fermion TFIM. At $f=1/2$, the mean values are approximatly 
$4.87$, $4.81$, and $4.81$ for the integrable,
chaotic, and random-state cases, respectively.
Their standard deviations are $0.0843$, $7.46\times 10^{-4}$,
and $3.28\times 10^{-4}$.
The chaotic value closely agrees with the random-state
reference, whereas the integrable chain has a higher value and much larger eigenstate-to-eigenstate fluctuations.
At $f=1/2$, the integrable mean remains above the random-state
reference through $N=24$ [Fig.~\ref{fig:xxz_scaling}].

\begin{samepage}
For the interacting integrable XXZ chain, our numerical results also indicate that the stabilizer
correlation varies weakly with $f$ near an equal bipartition,
although its value differs from the
random-state prediction and its eigenstate-to-eigenstate
fluctuations are much larger. This behavior parallels the observation in
Ref.~\cite{sm:odavicquench} that the long-time behavior
of the stabilizer entropy in the XXZ chain resembles
that in nonintegrable systems. A more detailed
investigation of this behavior is left for future work.
\par\end{samepage}

\end{document}